\documentstyle[12pt,moriond,epsfig]{article}
\def\la{Ly$\alpha$}
\def\hb{H$\beta$}

\def\kms                 {km\thinspace s$^{-1}$}

\def\aa{{\rm A$\,$\&$\,$A}}            
\def\aas{{\rm A$\,$\&$\,$AS}}          
\def\apj{{\rm ApJ}}                    
\def\aj{{\rm AJ}}                      
\def\mnras{{\rm MNRAS}}                        

\def\Msun{\thinspace\hbox{$\hbox{M}_{\odot}$}}
\def\deg{\hbox{$^\circ$}}
\def\kpc{\thinspace\hbox{kpc}}

\begin{document}

\heading{ The Lyman-alpha emission in local Star-Forming Galaxies:
Scenario and Connection with Primeval Galaxies}

\author{D. Kunth $^{1}$,  E. Terlevich $^{2}$, R. Terlevich $^{3}$\footnotetext[3]{Visiting Prof. at INAOE}, G. Tenorio-Tagle $^{2}$} {$^{1}$Institut d'Astrophysique de Paris, 98bis Bld Arago, F-75014 Paris,
 France } {$^{2}$ INAOE,Tonantzintla, Apartado Postal 51 y 216, 72000 Puebla,
 Mexico } {$^{3}$ Royal Greenwich Observatory, Madingley  Road, Cambridge CB3 0EZ, UK} \\

\begin{moriondabstract}
We review the \la\ emission in local star-forming galaxies. In most cases as
 already shown by the IUE, the emission
 is absent or much weaker than expected. This occurs because \la\ photons can 
be resonantly scattered by the neutral gas and destroyed by even very low 
amounts of dust. 
However new Hubble Space Telescope observations (HST) indicate that
 other factors  such as the velocity 
structure of the gas play a crucial role. Gas flows are likely to occur
 as powered by the kinetic
energy released via stellar winds and supernova. We propose a scenario based
on the hydrodynamics of superbubbles powered by massive bursts of star
 formation that naturally accounts for the variety of \la\ line detections
 in star-forming galaxies. We caution with the attempts to derive the
 co-moving star formation rate at high redshift from \la\ emission searches.

\end{moriondabstract}

\section{Introduction}

The search for high-redshift galaxies have progressed rapidly over the last 
ten years using color techniques which basically select galaxies with massive
ongoing star formation and  little extinction while more evolved objects
that could be heavily dust-reddened are picked out at millimeter wavelengths
(see \cite{HC} for references).  It has been conjectured that primeval
 galaxies at their very early stage would be nearly dust-free hence
easily detected from their \la\ emission (\cite{PP}; 
\cite{M}). For this reason it is reasonable to speculate that
local star-forming 
galaxies  producing \la\ emission  would ressemble
distant primeval ones. 

Early ultraviolet observations of nearby starburst galaxies however, 
have 
revealed in most starburst galaxies a  much weaker \la\ emission than
 predicted by simple models 
of galaxy formation. In some other galaxies \la\ was non-existent or even 
appeared as a broad absorption profile (\cite{MT},
\cite{H84,H88}; \cite{DJK}; 
\cite{TDT}; \cite{K97,K98}). The reason for the
weakness of the \la\ emission has been the subject  of debates.
 For young star-forming galaxies without so much dust
as to  suppress \la\,  large equivalent widths should be observed in the
 range of
100-200 (1+z)~\AA \ \cite{CF}. 
However it was early realized that pure extinction by dust would be unable to
explain the low observed \la /\hb\ , although Calzetti and Kinney 
 \cite{CKI} tentatively proposed 
that proper extinction laws would
correctly match the predicted recombination value. 
Valls-Gabaud \cite{VG} on
the other hand suggested that ageing  starbursts could reduce 
\la\ equivalent widths because they are affected
by strong underlying  stellar atmospheric absorptions.  Early IUE data have 
provided evidence for an anticorrelation between the  \la /\hb\ ratio and
 the HII galaxy metallicity. These results were
attributed to the effect of resonant scattering of \la\ photons
and their subsequently increased absorption by dust (\cite{MT}; \cite{CF} 
and references therein).
Indeed, the enormous increase in optical path length experienced by the \la\
photons implies that small amounts of dust are able to completely destroy
the emission line, even originating a broad, damped absorption profile 
 \cite{CN}.\\
Finally Charlot and Fall \cite{CF} advocated that
the structure of the interstellar medium (porosity and multi-phase structure) 
is most probably an important factor for the visibility of the \la \ line.

\section{HST observations}

\begin{figure} 
\begin{center}\mbox{\epsfxsize=15cm \epsfbox{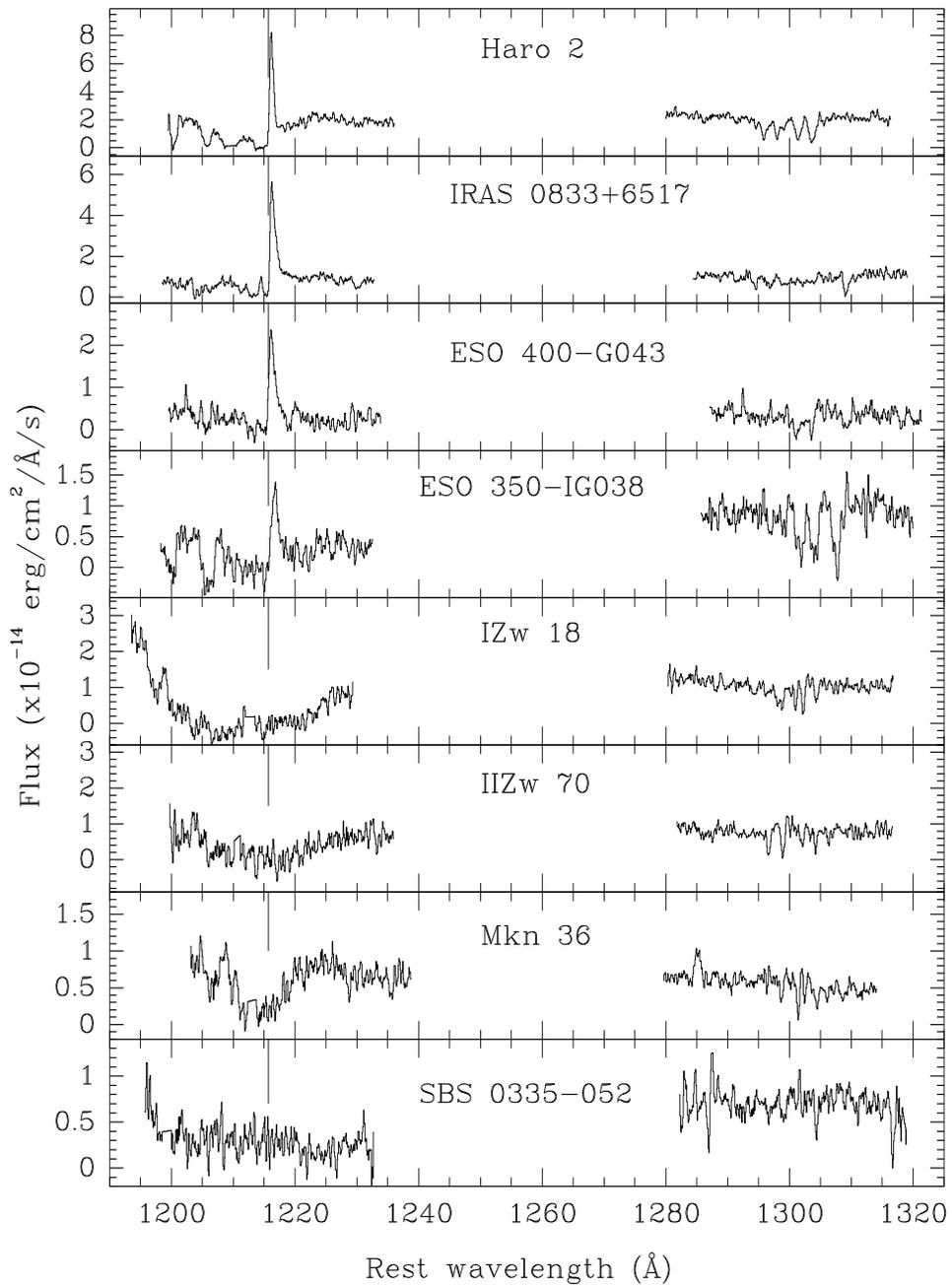}}\end{center}
\caption[]{
GHRS spectra of all the galaxies in \cite{K98}. The spectra have been
shifted to rest velocity assuming the redshift derived from optical
emission lines. Vertical bars indicate the wavelength at which the Ly$\alpha$\
emission line should be located. The geocoronal emission profile has been
truncated for the sake of clarity. The spectra have been plotted after
rebinning to 0.1~\AA\ per pixel and smoothed by a 3 pixel box filter.  
}
\label{fig:total}
\end{figure}

New  observations performed with the HST bring additional insight into this
picture. They
indicate that the velocity structure in the
interstellar medium plays a key role in the transfer and escape
 of Ly$\alpha$\
photons. Kunth et al. \cite{K94}
and Lequeux et al. \cite{LKM} have used the Goddard High Resolution Spectrograph
(GHRS) onboard the Hubble Space Telescope (HST).The Large Science
Aperture  was chosen to ensure a
sufficient flux level and  grating angles were selected according to the
redshift of the objects, so as to cover the Ly$\alpha$\ and the
 O\,{\sc I} 1302.2~\AA
\ and Si\,{\sc II} 1304.4~\AA\
regions.  The Ly$\alpha$\ range was chosen to
investigate both emission and absorption features so that the H\,{\sc I}
 column
density could be estimated.  The O\,{\sc I} 1302~\AA\ and 
Si\,{\sc II} 1304~\AA\ region was
selected to crudely estimate the chemical composition of the gas and to
measure with reasonable accuracy the mean velocity at which the absorbing
material lies with respect to the star-forming region of a given galaxy.
In a first attempt, Ly$\alpha$\ was observed only in absorption in the starburst
dwarf galaxy IZw~18, \cite{K94} . Since IZw~18 at $Z=1/50
~Z_{\odot}$ is the most metal--poor starburst galaxy known at present, it
was considered previously a good candidate to show Ly$\alpha$\ in emission.
 To add to the confusion, a  Ly$\alpha$\ emission line showing a complicated
profile, but a clear P~Cygni component, has been detected in Haro~2, a 
 dustier star-forming galaxy with a $Z=1/3 ~Z_{\odot}$ (\cite{LKM}).
The detection of such a profile in the
 Ly$\alpha$\
emission line of Haro~2 led to postulate that the line was visible
because the absorbing neutral gas was velocity--shifted with respect to the
ionized gas. This was confirmed by the analysis of the UV O\,{\sc I} and 
Si\,{\sc II}
absorption lines (blue--shifted by 200~km\thinspace
 s$^{-1}$\  with respect
 to the optical emission lines) and that of the profile of the H$\alpha$ line
(\cite{LKML}).
Observations were subsequently made on additional  galaxies  by Kunth et al.
\cite{K98} while Thuan and Isotov \cite{TI} have obtained GHRS 
spectra of two more starburst galaxies,
namely Tol65 and T1214-277  with the G140L grating allowing  a larger
 spectral region around \la \ . Tol65 reveals a broad damped \la\ absorption
 while T1214-277 
shows a pure \la\ emission profile with an equivalent width of 70 \AA \ 
and with no blue absorption. 
The individual spectra of galaxies observed in \cite{K98} are shown in
Fig.~\ref{fig:total}

\section{Results of HST observations and interpretation}

\subsection{The Local Star-Forming Galaxies}
 Three types of observed lines have been identified so far: pure \la\ 
emission;  broad damped  \la\ absorption  centered at the 
wavelength corresponding to the redshift of the HII emitting gas
  and \la\
emission with blue shifted absorption features, leading in some cases 
to P Cygni profiles.

\begin{figure} 
\begin{center}\mbox{\epsfxsize=9cm \epsfbox{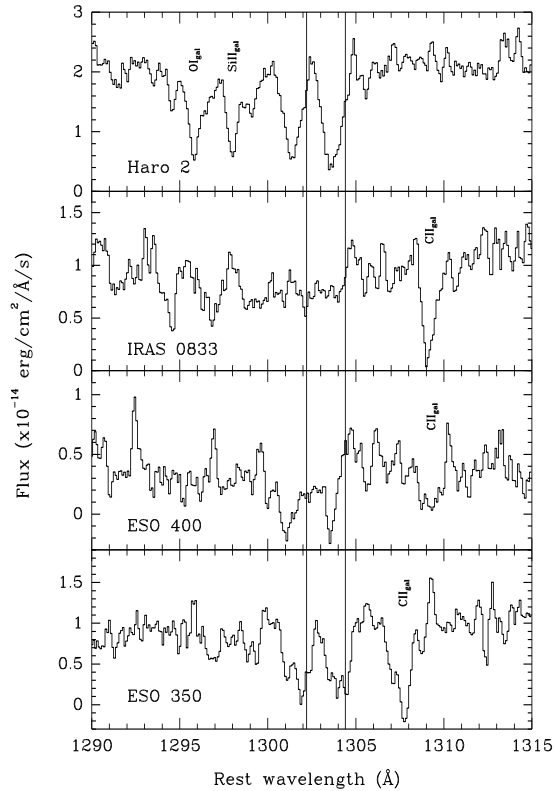}}\end{center}
\caption[]{
Detail of the O\,{\sc I} and Si\,{\sc II} region for the galaxies
 with Ly$\alpha$\
emission. The vertical bars indicate the wavelength at which the O\,{\sc I} and
Si\,{\sc II} absorption lines should be located, according to the redshift
 derived
from optical emission lines. Some Galactic absorption lines have been
marked. Note that the metallic lines appear systematically blueshifted in
these galaxies with respect to the systemic velocity. In some cases there
is no significant absorption at all at zero velocity. }
\label{fig:oila}
\end{figure}

\begin{figure} 
\begin{center}\mbox{\epsfxsize=9cm \epsfbox{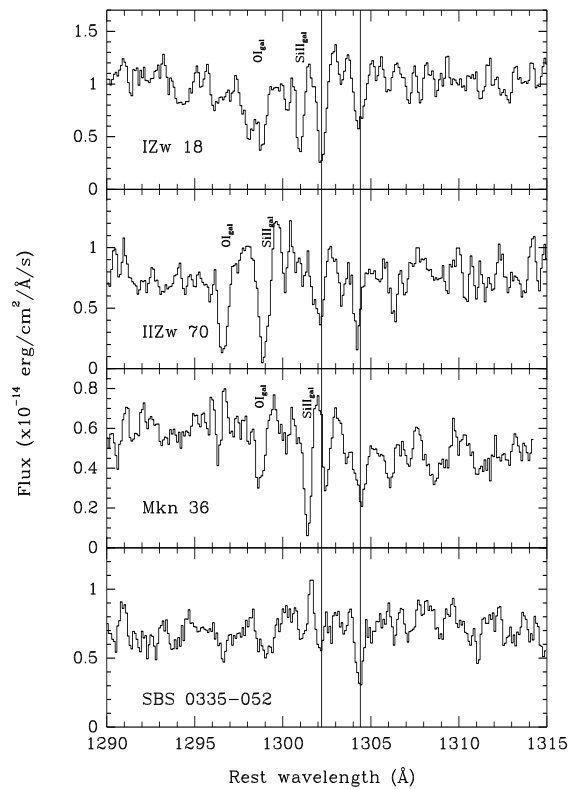}}\end{center}
\caption[]{
O\,{\sc I} and Si\,{\sc II} region for the galaxies showing damped Ly$\alpha$\
absorptions. Details as in Fig.~\ref{fig:oila}. Note that in these
galaxies the metallic lines are essentially at the same redshift than the
ionized gas, indicating the presence of static clouds of neutral gas, as
discussed in the text. }
\label{fig:oiab}
\end{figure}

As noted by \cite{K98},  \la\ emission with deep blueward absorption troughs
 evidence a wide velocity field. The equivalent widths
 of the \la\ emission range
between 10 and 37~\AA \ hence much below the value predicted
by Charlot and Fall, \cite{CF} for a dust free starburst model. 
In all cases interstellar absorption lines (OI, SiII) are
 significantly blueshifted with respect to the HII gas
 (see Fig.~\ref{fig:oila}).
On the other hand, if the HI is static with 
respect to HII, the destroyed \la\ 
photons are those emitted by the HII region and the interstellar
lines are not displaced (see Fig.~\ref{fig:oiab}). 

The details of the processes are as follows. 
Ly$\alpha$\ photons are produced by recombinations in H\,{\sc II} regions at about 
2/3 of the ionization rate. They are subsequently absorbed and reemitted by H
atoms, both in the H\,{\sc II} regions in which they were produced and in the
surrounding H\,{\sc I} regions, if present. This process -- resonant 
scattering -- changes both the frequency and direction of the Ly$\alpha$\ photons.
Therefore those produced within a galaxy would eventually escape from it,
in one direction or another.
 This scattering process increases enormously the mean free path
of the trapped photons, so that if some dust is present, the probability of
absorption around the Ly$\alpha$\ wavelength increases also by a significant
factor with respect to the standard UV extinction.  As a consequence,
absorption is potentially important whenever the dust-to-gas ratio exceeds
about one percent of the Galactic value (see, e.g., Eq. (3) of \cite{CF}).

A further complication arises if the neutral gas
surrounding the star-forming regions is not static with respect to the
ionized gas, but  outflows from these regions towards the observer. The
resonant scattering would affect photons at shorter wavelengths than the
Ly$\alpha$\ emission line, i.e.,  photons resonantly trapped, and potentially
destroyed by dust, would be mostly stellar continuum photons emitted at
wavelengths below 1216~\AA \ . For a galaxy in which the source of ionizing
radiation is a stellar population with a normal initial mass function, the
angle-averaged equivalent width of the Ly$\alpha$\ emission line is about 100~\AA\
in the dust-free case (\cite{CF}). This depends only weakly on
the star formation rate in the galaxy provided it is reasonably continuous
(and nonzero over the past few $\times10^7$~yr).  This value can be
somewhat higher if instead the star formation episode is ``instantaneous'',
i.e. if it lasts less than a few $\times10^6$~yr, as it seems to be the case in
most compact star-forming galaxies. Nevertheless, since the Ly$\alpha$\ photons
would diffuse (in the dust-free case) through the external surface of the
neutral clouds (which are rather large in these compact star-forming
galaxies, extending far beyond the optical regions), its surface brightness
would be very small.  Therefore, even in a dust-free case, we would expect
to detect an absorption line around the Ly$\alpha$\ wavelength if the aperture
sustended by the slit is small compared to the spatial extension of the
neutral cloud. This absorption will be centered at the wavelength
corresponding to the mean velocity of this neutral gas, i.e., it will be
blueshifted with respect to the Ly$\alpha$\ emission line if the neutral gas is
moving towards the observer.\\  
 Among the 
galaxies showing a strong damped Ly$\alpha$\ absorption at the systemic velocity,
the O\,{\sc I} and 
Si\,{\sc II}
appear in absorption without any significant
velocity shift with respect to the H\,{\sc II} regions. This indicates that the
neutral gas in which they mostly originate is static with respect to the
star-forming region. Therefore, since these galaxies have a low dust
content - IZw~18 shows weak signs of reddening
and its dust-to-gas ratio is at least 50 times smaller than the Galactic
value, \cite{K94} -, this shows that it remains possible to weaken
observationally  Ly$\alpha$\ by simple multiple resonant scattering from the
neutral gas, and even to produce an absorption feature. The H\,{\sc I} cloud 
surrounding these galaxies might be leaking Ly$\alpha$\
photons through its external surface. The Ly$\alpha$\ line would then become very
hard to detect because of its low surface brightness. This extended
emission could be detected with deep, large area observations around these
galaxies. Nevertheless, it might be that even the small amount of dust
present in these galaxies   efficiently destroys a significant
fraction of Ly$\alpha$\ photons, especially if the clouds extension is very
large. \\
On the other hand, the Ly$\alpha$\ emission in Haro~2 is accompanied by a broad
absorption in the blue wing of the line, with the general appearance of a
typical P~Cygni profile. The amount of neutral gas that produces the blue
absorption trough at Ly$\alpha$\ is rather modest and of the order of
N~(H\,{\sc I}) = 7.7x$10^{19}$ atoms cm$^{-2}$. The crucial point here
 is that the
neutral gas responsible for the absorption  is not at the
velocity at which the Ly$\alpha$\ photons were emitted. Moreover, it seems that
all the neutral gas along the line of sight is being pushed by an expanding
envelope around the H\,{\sc II} region, outflowing at velocities close to 200
km\thinspace s$^{-1}$. This interpretation is of course strengthened by
 the presence of
other detected absorptions of O\,{\sc I}, Si\,{\sc II} and Si\,{\sc III}
 due to 
outflowing gas in
front of the ionizing hot stars of the central H\,{\sc II} region. 
 Legrand et al., \cite{LKML} 
have obtained high resolution spectroscopic observations of Haro 2 in
 H$\alpha$ with
the WHT at La Palma, finding evidence for an
expanding shell.
Comparison of the Ly$\alpha$\ and the H$\alpha$ profiles shows that the
 Ly$\alpha$\ line
is significantly broader than H$\alpha$, suggesting also scattering of
photons from the back side of the expanding neutral cloud.\\
Data on other  H\,{\sc II} galaxies with detected Ly$\alpha$\ emission
confirm that Haro 2 is not an isolated case. Most spectra show Ly$\alpha$\ emission
with a broad absorption on their blue side except for ESO 350-IG038 in
which the emission is seen atop of a broad structure requiring several
filaments and T1214-277 that shows pure emission.
 When  metallic lines are detected, they are always
blueshifted with respect to the ionized gas, further supporting the
interpretation. In the case of ESO 350-IG038 the velocity structure seems
to be more complicated and several components at different velocities are
identified on the metallic lines. Note that there might be a secondary peak 
emission in the
 blue side of the main line in the spectrum of ESO-B400-G043.

 Ly$\alpha$\ absorption profile fitting requires one or several components (in
addition to a Galactic component if the redshift is small) with
 relatively little scatter in the derived column densities.
  Most clouds have a column density log~N(H\,{\sc I}) of
nearly 19.7 to 21.5. The static clouds tend to have larger column densities
than the moving ones that are also split into several components as
expected in a dynamical medium.

The main conclusion drawn from this set of data is that complex
velocity structures are determining the Ly$\alpha$\ emission line 
detectability,
showing the strong energetic impact of the star-forming regions onto their
surrounding ISM. This velocity structure is therefore the driving factor for
the Ly$\alpha$\ line visibility in the objects of our sample.  We want to stress
that this effect seems to be almost independent of
the dust and metal abundance of the gas. 

If the neutral gas is static with respect to the H\,{\sc II} region, the 
covering
factor by these neutral clouds would probably become the key factor
determining the visibility of the line. Thuan and Isotov \cite{TI} have 
detected strong Ly$\alpha$\ emission in T1214-277, with no evidence of
blueshifted Ly$\alpha$\ absorption. In this case the detection of the line
requires that a significant fraction of the area covered by the slit along
the line of sight is essentially free from neutral gas, suggesting a patchy
or filamentary structure of the neutral clouds. Such a geometry would be
unlikely  in galaxies surrounded by enormous H\,{\sc I} clouds, as 
in IZw~18 and similar objects. 

\subsection{The Galaxies at High-Redshift}

The effect of neutral gas flows helps to understand why luminous
high-redshift objects have only been found up to now with linewidths
larger than
1000 km s$^{-1}$.  High--redshift galaxies with very strong (EWs $>$
500~\AA) extended Ly$\alpha$\ emission are characterized by strong velocity shears
and turbulence (v $>$ 1000 km s$^{-1}$); this suggests an AGN activity,
in the sense that other ISM energising mechanism than photoionization by
young stars may be operating. However Steidel et al. \cite{SGP} have recently
discovered a substantial population of star--forming galaxies at
3.0$<$z$<$3.5 that were selected not from their emission--line properties
but from the presence of a very blue far-UV continuum and a break below
912~\AA\ in the rest frame. Similarly to our local starbursts they find that
50\% of their objects show no Ly$\alpha$\ emission whereas the rest does, but with
weak EWs no larger than 20~\AA\ at rest.  The Ly$\alpha$\ profiles of this
population look indeed very similar to those of our local starburst
galaxies (\cite{F} ; \cite{PS}; see also \cite{DSS} for the z=5.34 galaxy).
We can conclude from the preceeding discussion that the use
 of Ly$\alpha$\ as a star
formation indicator underestimates the comoving star formation density
at high redshift (\cite{HC}).

\section{The evolution of superbubbles in extended HI halos}

\begin{figure} [p]
\begin{center}\mbox{\epsfxsize=15cm \epsfbox{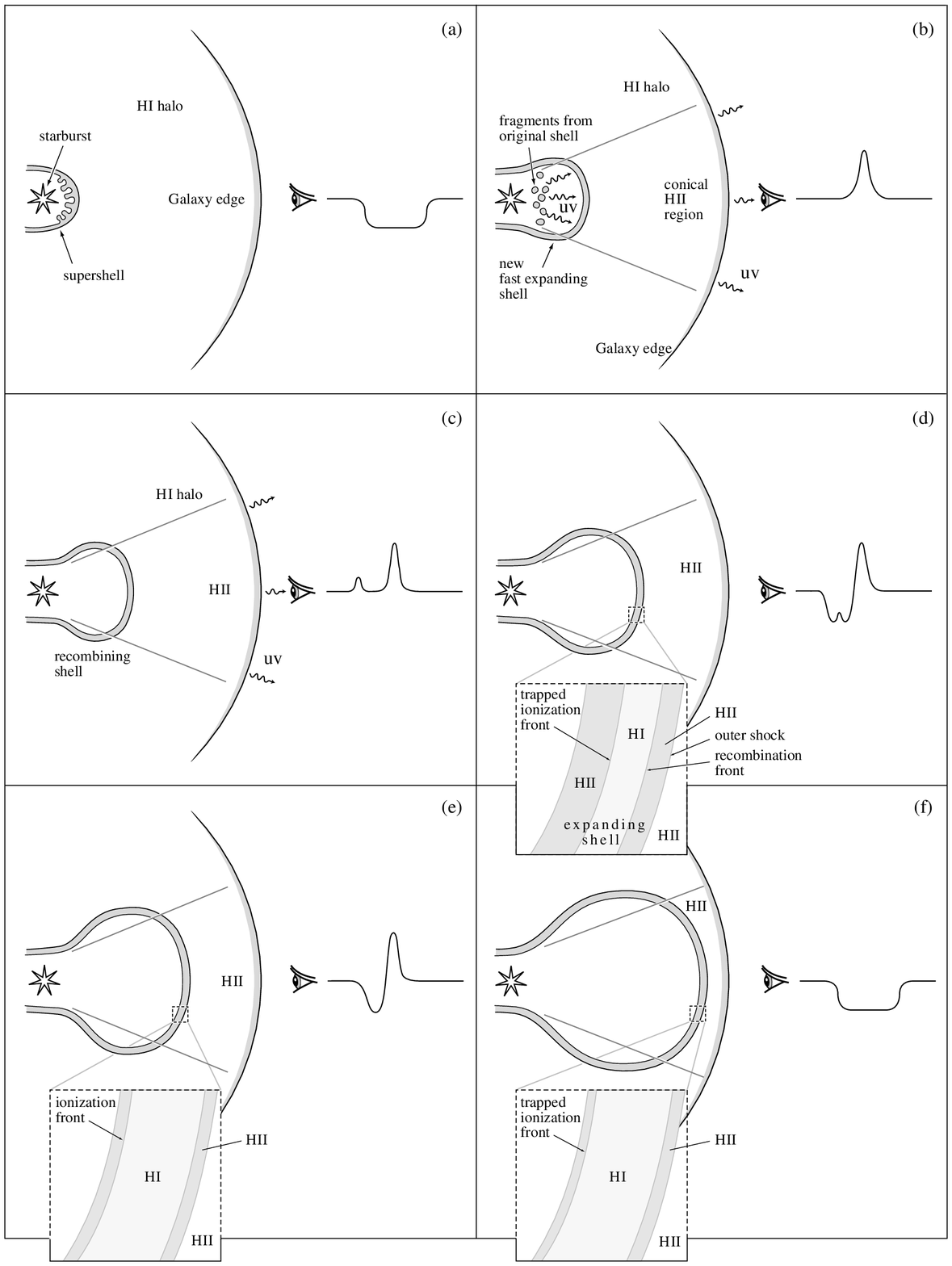}}\end{center}
\label{fig:superbubble}
\end{figure}

To account naturally  for the variety of 
\la\ line detections in star-forming galaxies Tenorio-Tagle et al. \cite{TKT},
have proposed a scenario based on the hydrodynamics of superbubbles powered by
 massive starbursts. This scenario is  visually depicted in
Fig.~\ref{fig:superbubble}.
The overpowering mechanical luminosity ($E_0$ $\geq$ 10$^{41}$erg s$^{-1}$)
from massive starbursts ($M_{stars}$ = 10$^5$ - 10$^6$ \Msun) is known 
to lead to a rapidly evolving superbubble able to blowout the gaseous 
central disk configuration into their extended HI haloes, 
as $E_0$ exceeds the energy input rate required 
for the remnant to reach a disk scale-height with supersonic velocity. As shown 
by \cite{KK} if $E_0$ exceeds the threshold luminosity 
$L_b$ = 7.2 $ \times 10^{36} P_4 H_{100}^2 a_{s,10}$  erg s$^{-1}$
(where $P_4$ = disk pressure in units of $(10^4k)$ cm$^{-3}$ K, $H_{100}$ is the
 disk scale height in
units of 100 pc and $a_{s,10}$ is the disk sound speed in units of 
10 \kms), the superbubble will blowout and thus massive starbursts 
will lead to  superbubble blowout phenomena even in massive galaxies 
such as the Milky Way. One can then predict  the venting of the  
hot superbubble interior gas through the 
Rayleigh-Taylor fragmented shell, into the extended HI halo
where it would  push once again the outer shock allowing it to build a new shell
of swept up halo matter (see Fig.~\ref{fig:superbubble}b). 
Blowout occurs quite early in the evolution of the starburst ($T_{blowout} \leq $ 
2Myr).

A coeval starburst with a total mass larger than several
$10^5$\Msun , also produces an almost constant ionizing photon flux 
($F_{uv} \geq 10^{52} $ photons s$^{-1}$) during the 
first few (3-4) Myr and then abruptly, after $t$ = $t_{ms}$, it
begins to decrease as t$^{-5}$ (see \cite{BTY}, \cite{MHK}) 
as the most massive stars begin to 
evolve away from the main sequence to eventually end up as supernovae. 
But then blowout or the fragmentation of the shell that allows
the expansion of the hot interior gas into the extended low density HI halo, 
also allows the leakage of the $uv$ photons emitted by the starburst.
These photons establish an ionized conical 
HII region with its apex at the starburst. The density of the halo
steadily decreases with radius, but even if one regards it as a constant 
density medium with an $n_{halo}$ $\sim$ 10$^{-2}$cm$^{-3}$, one 
can show that the conical sector of the HII region will extend all the way to 
the galaxy outer edge, i.e.~several \kpc. This is because the 
isotropic stellar radiation manages to support along the plane of the 
galaxy  an ionized central region of say, typically, 100 pc in radius,  
at the very core of the giant HII region where the density is the largest
($n_{core}$ $\sim$ 1 - 10 cm$^{-3}$). Thus, along the breakout direction 
it could support the ionization of a region of length 
$l$ = 100pc ($n_{core}/n_{halo})^2$, several \kpc\ long. Furthermore, 
the low halo density implies a
long recombination time  (t$_{rec}$ = 1/($\beta n_{halo}$) 
$\geq$ 10$^{7}$ yr), 
with the implication that
the ionized conical sector of the galaxy halo becomes, and remains,  
transparent to the 
ionizing flux produced  during the ensuing lifetime of the 
ionizing massive stars ($\leq$ 10 Myr). Then clearly, the \la\ photons
produced at the central HII region will also be able to travel freely 
in such directions. The escape of $uv$ photons from the 
galaxy would be particularly important during the early stages after
blowout
and until the new shell of swept up halo matter condenses enough 
material as to allow its recombination. This will be favoured if the  
shock progresses with speeds of a 
few hundred ( $\leq$ 400)
km s$^{-1}$, (i.e. with a Mach number $M$ $\leq$ 40)
leading to temperatures: $T_{shock} = 
1.4 \times 10^7 (v_{shock}/1000$ km s$^{-1}$)$^2$, near the top of the 
interstellar cooling curve. This fact will promote the rapid cooling 
of the shocked gas in a time $t_{cool} = 3 k T_{shock}/(4 \Lambda n_{halo})$,
 where $\Lambda $ is the interstellar cooling rate. 
Given the values of $\Lambda$ ($\sim$ a few times 10$^{-22}$ erg cm$^3$ s$^{-1}$)
 and of $n_{halo}$, $t_{cool}$ 
is of the order of one Myr; but note that lower density haloes 
(with an $n_{halo} \sim 10^{-3}$) will not have time to cool within 
the life-time of the massive stars. Rapid cooling 
implies that the shocked gas will cool down to the HII region temperature 
making the shock isothermal, and thus causing  compression
factors of several hundreds (as  
the shocked gas density $n_{shock}=n_{halo}M^2$).   
Compression leads then to recombination in time-scales 
($t_{rec} =  1/\beta n_{shock})$ of less than 
10$^5$yr,  and this immediately
and steadily will reduce the number of stellar $uv$ photons leaking out of the 
galaxy. At the same time recombination in the fast expanding shell 
will lead to a correspondingly blueshifted \la\ emission, as depicted 
in Fig.~\ref{fig:superbubble}c.
Once the shell presents a large column density 
($\sim$ 10$^{19}$ atoms cm$^{-2}$), as it grows to dimensions of a few kpc, 
it will trap the  
ionization front. This is promoted by the large shell densities and
the geometrical dilution of the ionizing radiation. Note that 
from then onwards,  
recombinations in the shell  will inhibit the 
further escape of ionizing photons from the galaxy (compare
 Fig.~\ref{fig:superbubble}b,
 c, 
and d).
The trapping of the ionization front,  makes the shell, 
acquire a multiple structure with a photoionized inner edge, 
a steadily growing central zone of HI, and an outer ionized sector where 
the recently shocked 
ionized halo gas is steadily incorporated. 
The growth of the central layer eventually will cause  
sufficient scattering and absorption of the \la\ photons 
emitted by the central HII region, leading to a blueshifted \la\ 
absorption. 
Note also that for as long as recombinations continue to occur at the 
leading edge of the shell, a blueshifted \la\ in emission will appear 
superposed to the 
blueshifted absorption feature (see Fig.~\ref{fig:superbubble}d). Recombination at the 
leading edge of the shell will become steadily less frequent, depleting the 
blueshifted \la\ in emission. This is due when the shell and its 
leading shock move into 
the outer less dense regions of the halo, and the shell recombination time, 
despite the compression at the shock, becomes larger than the dynamical time. 
At this stage, an observer looking along the conical sector of the HII region
will detect a P-Cygni-like \la\ line profile as shown in Fig.~\ref{fig:superbubble}e. 

Geometrical dilution 
of the $uv$ flux will begin to make an impact as the superbubble grows 
large. This 
and the drop in the $uv$ photon production rate, 
caused by the death of the most massive stars after  
$t$ = $t_{ms}$, will enhance the
column density of neutral material in the central zone of the 
recombined shell to eventually cause the full saturated 
absorption of the \la\ line (see Fig.~\ref{fig:superbubble}f).
 Full saturated absorption has usually been accounted for by the large
column 
density of the extended HI envelope of these galaxies and thus, as in all 
models, many different orientations will suit the observations. In our
scenario however, also when  
observing within the solid angle defined by the conical HII region formed 
after blowout, such  broad absorptions could arise well after
blowout, 
once a large column density of shocked neutral material ($N$ $ \geq$ 
10$^{20}$ atoms cm$^{-2}$) has formed ahead of the trapped ionization front.

The solid angle subtended by the ionized cone will be a rapidly changing 
function of time, particularly during the early stages of evolution, 
immediately after blowout. However, numerical experiments and observations
(see \cite{ST}, and \cite{TM}
and references therein) restrict this to a 
maximum value of about 70\deg , with the walls of the superbubble
near the galaxy plane inhibiting its further growth. 

\section{Discussion} 
The main implication of the evolution depicted in Fig.~\ref{fig:superbubble}
is that it is the feedback from the 
massive stars what -- through ionization  and the evolution of superbubbles --
leads to  the large variety of \la\ emission profiles. 
The escape of \la\ photons  
depends sensitively on the column density of the neutral gas and dust 
following the suggestion  that the attenuation by
dust is enhanced by scattering with hydrogen atoms.
Note that apart from 
the observed profiles: \la\ in emission, P-Cygni-like profiles and 
full saturated absorption (as in Fig.~\ref{fig:superbubble} a, b, e and f) the scenario 
predicts also secondary blueshifted \la\ emission profiles emanating 
from the rapidly expanding and recombining shell 
(see Fig.~\ref{fig:superbubble}c and d). If massive star formation leads also
 to networks of
shells such as those observed in 30 Dor (\cite{CK}) one 
should also expect a forest of \la\ in emission arising from 
recombinations in the various expanding shells in the network.      
P-Cygni \la\ profiles are predicted when 
observing along the angle subtended by the conical HII region but only once 
the ionization front is trapped by the sector of the superbubble shell
evolving into the extended halo. This will produce the fast moving 
layer of HI at the 
superbubble shell, here thought to be responsible for 
the partial absorption observed in sources such as Haro 2, ESO B400-G043 (which
probably exhibits a secondary blueshifted \la\ emission) and
 ESO 350-IG038 (\cite{K98}). In the latter case however, the profile
is not typical of a clear P Cygni one. Instead the underlying  damped \la\
 absorption extends beyond the red of the line emission. 

Damped \la\ absorption is seen in several galaxies. We note that 
these objects are all gas rich dwarf galaxies whereas in most cases but 
Haro 2, the galaxies that exhibit \la\ in emission or with a P-Cygni profile,
are on the higher luminosity side of the distribution ($M$ $\leq$ $-18$).

Pure \la\ emission is observed in C0840+1201 and T1247-232 
(\cite{TDT}; IUE)  or T1214-277 (\cite{TI}; HST). 
Such a line implies no absorption and thus no HI gas between the starburst 
HII region and the observer, as when observing the 
central HII region after the superbubble blowout, within the conical HII
region carved in the extended HI halo. 
It is not a straigthforward issue to estimate what is the 
fraction of Lyman continuum 
radiation that leaks out from galaxies.
Direct observations  from nearby star-forming 
galaxies indicate that less than 3\% of the intrinsic Lyman continuum photons 
escape into the intergalactic medium (\cite{LF}) while a more
restrictive value of less than 1\% has been obtained using the H$\alpha$ 
luminosity density of nearby galaxies (\cite{DF}). Our scenario 
however predicts a short but significant evolutionary 
phase (between blowout and the trapping of the ionization front 
by the fast expanding shell) during which a large amount of  $uv$ 
radiation 
could leak out of a galaxy into the intergalactic medium. 
Detailed numerical calculations of the scenario proposed here are
currently underway. Our results and further implications of the model 
will be reported in a forthcoming communication.

\vfill
\end{document}